# Ultrafast time-evolution of chiral Néel magnetic domain walls probed by circular dichroism in x-ray resonant magnetic scattering.

Cyril Léveillé[1], Erick Burgos-Parra[1,2], Yanis Sassi[2], Fernando Ajejas[2], Valentin Chardonnet[3], Emanuele Pedersoli[4], Flavio Capotondi[4], Giovanni De Ninno[4,5], Francesco Maccherozzi[6], Sarnjeet Dhesi[6], David M. Burn[6], Gerrit van der Laan[6], Oliver S. Latcham[7], Andrey V. Shytov[7], Volodymyr V. Kruglyak[7], Emmanuelle Jal[3], Vincent Cros[2], Jean-Yves Chauleau[8], Nicolas Reyren[2], Michel Viret[8] and Nicolas Jaouen[1]

[1] *Synchrotron SOLEIL, Saint-Aubin, Boite Postale 48, 91192 Gif-sur-Yvette Cedex, France*

[2] *Unité Mixte de Physique, CNRS, Thales, Université Paris-Saclay, 91767 Palaiseau, France*

[3] *Sorbonne Université, CNRS, Laboratoire Chimie Physique – Matière et Rayonnement, LCPMR, 75005 Paris, France*

[4] *Elettra-Sincrotrone Trieste, 34149 Basovizza, Trieste, Italy*

[5] *University of Nova Gorica, 5000 Nova Gorica, Slovenia*

[6] *Diamond Light Source, Didcot OX11 0DE, United Kingdom.*

[7] *University of Exeter, Stocker road, Exeter, EX4 4QL, United Kingdom.*

[8] *SPEC, CEA, CNRS, Université Paris-Saclay, 91191 Gif-sur-Yvette, France*

**Non-collinear spin textures in ferromagnetic ultrathin films are attracting a renewed interest fueled by possible fine engineering of several magnetic interactions, notably the interfacial Dzyaloshinskii-Moriya interaction. This allows the stabilization of complex chiral spin textures such as chiral magnetic domain walls (DWs), spin spirals, and magnetic skyrmions. We report here on the ultrafast behavior of chiral DWs after optical pumping in perpendicularly magnetized asymmetric multilayers, probed using time-resolved circular dichroism in x-ray resonant magnetic scattering (CD-XRMS). We observe a picosecond transient reduction of the CD-XRMS, which is attributed to the spin current-induced coherent and incoherent torques within the continuously dependent spin texture of the DWs. We argue that a specific demagnetization of the inner structure of the DW induces a flow of spins from the interior of the neighboring magnetic domains. We identify this time-varying change of the DW texture shortly after the laser pulse as**

**a distortion of the homochiral Néel shape toward a transient mixed Bloch-Néel-Bloch texture along a direction transverse to the DW.**

Ultrafast demagnetization of a ferromagnet by an optical pulse was first demonstrated in 1996 in the seminal study by Beaurepaire et al [Beaurepaire96], which is widely considered as the birth of the research field of femtomagnetism, *i.e.*, the magnetism modulated ("pumped") by femtosecond long laser pulses. While several underlying mechanisms are considered to explain these ultrafast processes, the central role of spin dependent transport of hot electrons has been clearly evidenced [Melnikov11, Siegrist19]. Such phenomena were first experimentally demonstrated in spin valves, in which the demagnetization process is faster for antiparallel alignment of the magnetization in the magnetic layers [Malinowski08]. Models based on polarized electron transport in the superdiffusive regime have been subsequently developed [Battiato10]. The optically excited hot electrons, initially ballistic, with spin-dependent lifetimes and velocities, generate non-equilibrium spin currents either within a ferromagnetic layer or in adjacent non-magnetic layer. The induced loss of angular momentum greatly participates in ultrafast dynamical behavior of the magnetization [vodungbo16]. The existence of this phenomenon has also been tested in single magnetic layers with a heterogeneous magnetization configuration, i.e., containing a large density of magnetic domains and DWs, albeit with different conclusions [Moisan14, vodungbo16, Pfau2012]. X-ray diffraction experiments are in this latter case more powerful for probing the behavior of DWs [zusin2020, Kerber20, Hennes20b]. For example, Pfau *et al*. [Pfau2012] inferred that the DW size changes in the first few ps by investigating the variations of the first-order Bragg peak of the magnetic configuration. More recently, the studies of Zuzin *et al*. [Zuzin2020] and Hennes *et al*. [hennes2020b] have shown that a more precise way to extract insights about DWs is to study the position and width of higher order diffraction peaks.

In this Letter, we use circular dichroism in x-ray resonant magnetic scattering (CD-XRMS) to gain access to the internal spin texture of the domain walls. This technique permits indeed a direct determination of the type (Néel or Bloch) as well as of the effective chirality of the DWs [Dürr99, chauleau2018]. Magnetic multilayers with homochiral Néel DWs stabilized by a large interfacial Dzyaloshinskii-Moriya (DM) interaction [Fert80, Fert90] are ideal systems to study DW dynamics at the fs timescales. In recent studies, this approach was used [Zhang17, chauleau2018, Legrand2018, Zhang20] to investigate the intrinsic nature of DWs and skyrmionic systems, which is currently a topic of the utmost relevance from both fundamental and technological viewpoints [Thiaville12, Ruy13, Nagoasa13, Fert17, Yang15]. The degree of circular dichroism in these experiments is not only related to the homochiral nature of the magnetic textures but also to the intrinsic DW configuration and allows us to probe the size and magnetization ratio of domain/domain-wall with unprecedented sensitivity. We hence unveil the ultrafast dynamics of these domain walls, unambiguously showing a specific behavior compared to that of the domains.

The system under study is an asymmetric magnetic multilayer [Pt(3 nm)|Co(1.5 nm)|Al(1.4 nm)]$_{x5}$ grown by sputtering on a thermally oxidized Si wafer buffered by Ta(5)|Pt(5) (see Supplementary Sec. S1 for details) presenting perpendicular magnetic anisotropy and large interfacial DM interaction. At remanence, domains adopt a typical disordered labyrinthine structure, but with a narrow distribution of domain widths. The magnetization and anisotropy are measured by SQUID magnetometry, while the DM amplitude is determined by comparing the experimentally measured (by magnetic force microscopy) domain periodicity to those simulated using micromagnetic calculations with MuMax3 [Vansteenkiste14] (see Supplemental Material S1 for details about the magnetic preparation and the simulations). From these calculations, we can also estimate the DW width to be ~20 nm. The micromagnetic simulations are also used as inputs in the empirical XRMS model with accurate values for the width of the DW.

The time-resolved XRMS experiments have been performed on the DiProI beam line [Capotondi13] at the FERMI free electron laser [Allaria12] (Trieste, Italy). Time resolution is achieved using a standard pump-probe approach [Fig. 1(a)] in which the probe is a 60 fs XUV pulse at the Co $M$ edge energy (photon energy ~60 eV) and the pump is a 100 fs infrared laser pulse (780 nm). The overall time resolution is therefore ~120 fs. The scattering experiments have been conducted under reflectivity condition at 45° incidence for circularly left (CL) and right (CR) x-ray polarization allowing to acquire ultrafast snapshots of diffraction diagrams (Fig. 1b) and their corresponding circular dichroism (Fig. 1c) at each delay time of the infrared (IR) excitation (see S2 for details). Noteworthy, the degree of x-ray circular polarization is between 92-95% [Allaria14]. Regarding the probe and pump energy densities, the IR fluence was set to 4.8 mJ/cm² (at a repetition rate of 50Hz) and the FEL fluence was set to 0.5 mJ/cm². At the Co $M$ edge, with 45° photon incidence angle, the penetration depth is ~10 nm, therefore most of the scattered signal comes from the uppermost Co layers. Such a small penetration depth also ensures that the expected tilting of the Ewald sphere is negligible in our experiment. Finally, we decided to perform the experiment at the peak of the absorption resonance to avoid any spurious effect caused by the energy shift of the XAS edge at ultrafast timescales [yao20, hennes20].

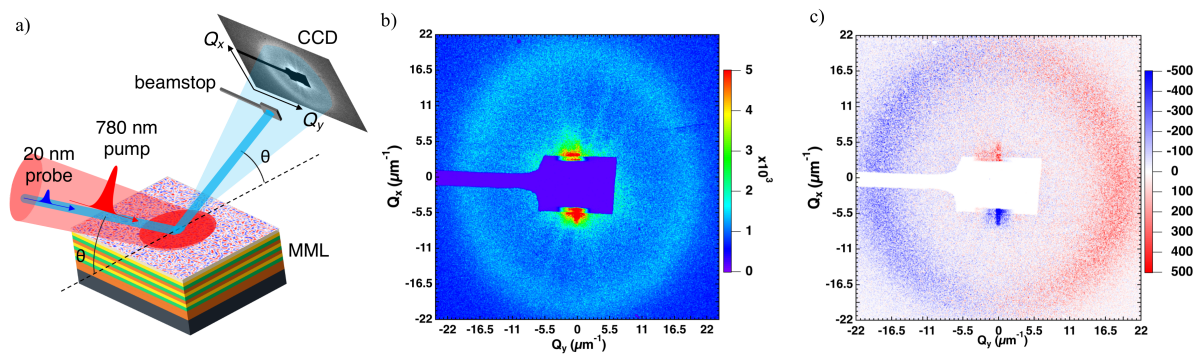

*Figure 1: CD-XRMS experiments. (a) Experimental configuration with the incident beams of the IR pump and the x-ray probe. (b) Magnetic diffraction pattern, (CL+CR) (c) Dichroic pattern (CL-CR), displaying the typical signature of clockwise Néel*

*domain walls. The images in panels b and c have been geometrically corrected to account for the projection related to the photon incidence angle θ = 45°, and the scale corresponds to the sum of the counts (500 XFEL pulse of each polarization) for (CL+CR) (b) and (CL-CR) (c).*

A typical diffraction pattern of the magnetic system at negative time delays, i.e. before the laser pulse excitation, is displayed in Fig. 1(b) in which the diffracted intensity is the sum of the two circular polarizations (CL+CR). It results from the x-ray diffraction on the labyrinth structure with a period of (330 ± 20) nm (estimated from the ring radius). The total magnetic scattering intensity mainly comes from the alternating out-of-plane magnetic domains. The diffraction intensity also displays circular dichroism (CL-CR) [Fig. 1(c)], which reverses its sign on each side (along $Q_y$) of the specular reflection, and reaches about 10%. Such dichroic signal is known [Durr99] to be a signature of an uncompensated sense of rotation in non-collinear magnetic textures. In our experiment, the sign of the dichroism indeed reveals the stabilization of clockwise (CW) Néel DW as we recently demonstrated [Chauleau2018]. The observed features have been corroborated by static scattering measurements at the Co *L* edge performed at the SEXTANTS beamline at SOLEIL [Sacchi13], for which the interpretation is now well established (see Supplementary Materials S1).

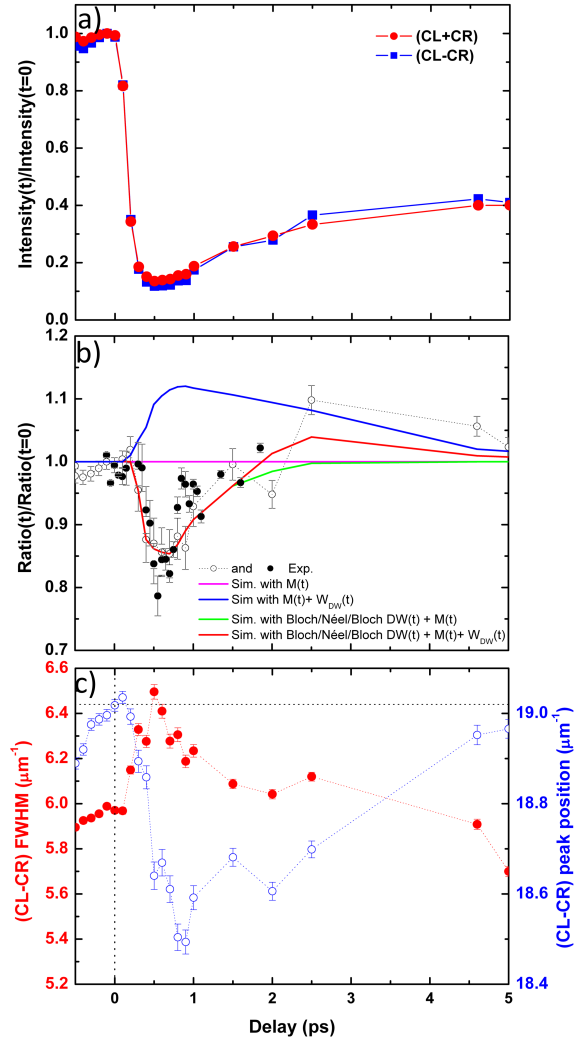

*Figure 2: Evolution of the XRMS signal over the first 5 ps: (a) intensity of integrated diffraction ring (CL+CR) and dichroism (CL-CR) normalized at their values at negative time delays; (b) experimental asymmetry ratio (CL-CR)/(CL+CR) normalized by its value at t < 0 in grey circles and black dots. The simulations for different models discussed in the main text appear as colored lines (see Supplementary Materials S3 for details). (c) Full width at half maximum (FWHM) (red dots) and the position (blue circles) in reciprocal space of the magnetic dichroic peak as a function of time.*

The time dependence of both the magnetic intensity (CL+CR) of the overall diffraction ring [Fig 2(a)] and the dichroism (CL-CR) shows a typical signature of ultrafast demagnetization in metallic magnetic ultrathin layers: first, a quench of the magnetization reaching a minimum value after a few hundreds of fs, followed by a log-like recovery over a few ps. The experimental results are further analyzed by plotting the asymmetry ratio, i.e., (CL-CR)/(CL+CR) as a function of time [see Fig. 2(b)], which represents the DW behavior normalized by the total magnetic moment. If the DW magnetization follows the same dynamics as that of the domains, this ratio should not vary. It is plotted in Fig. 2b (normalized by its value before the pump pulse) where one clearly observes a 15% dip at ~0.7 ps. This has been reproducibly observed when repeating the experiment, as demonstrated by the overlapping series of black filled and open circles in Fig. 2(b) showing identical behavior within error bars (inferred from the statistical fit of the background and peak intensity, see Supplemental Material Section S2). The normalized ratio remains below 1.0 up to 2 ps. The time evolution of the peak position defined by the maximum of its Gaussian fit, and of the full width at half maximum (FWHM) in reciprocal space of the magnetic dichroic peak are displayed in Fig. 2(c). Those two quantities generally correspond respectively to the variation of the domain size and their distribution. However, this *apparent* domain extension corresponds in fact to an expansion of the DW in the first ps after optical excitation, as reported by Pfau *et al*. [Pfau 2012]. When considering this expansion according to the value reported in Fig. 2(c), an *increase* of the asymmetry ratio is predicted as shown by the blue curve in Fig. 2(b).

To explain this ultrafast deviation of the dichroism asymmetry ratio, we first exclude an origin due to a change in the scattering factors induced by hot electrons filling the *d* band. Indeed, the IR laser fluence of our experiment is much lower (~10%) than the one used to probe the change of electron occupation induced by the IR pulse using x-ray absorption spectroscopy (XAS) [Mathieu18]. Thus, we explain our observation by the fact that during the demagnetization (resp. remagnetization), the magnetic moments do not decrease (resp. increase) by the same amount simultaneously inside the DWs and inside the domains. If the magnetization decreased uniformly, the expected asymmetry ratio would be constant, as shown by simulation using a model that is detailed in Supplemental Material S3 [magenta line in Fig. 2(b)]. As explained above, the sole expansion of the DW widths cannot explain our data [blue curve in Fig. 2(c)]. To explain an asymmetry ratio dropping below its initial value, we resort to a reduction of the degree of magnetic chirality. In other words, it corresponds to a change of the ratio between the out-of-plane and the in-plane magnetization. In our interpretation, the ultra-fast decrease of the asymmetry ratio below 1.0 is linked to a different demagnetization rate between the DWs and the domains. Note that a scenario that would correspond to a faster remagnetization of the DWs than the domains shall result into an asymmetry ratio larger than 1 (similarly to the expansion of the DW), and therefore can also be safely ruled out. In the following, in order to reproduce our experimental observations, the simulations include both coherent evolution of the hot electron spins that induce a spin torque on the DW and spin temperature (incoherent) variations within the DWs.

The understanding of the ultrafast DW width expansion requires considering the intense flow of spin currents in the ps regime. These can efficiently transfer angular momentum to and from the ferromagnetic material as shown, e.g., when Pt layers absorb it and generate ps electrical pulses [Kampfrath13]. Angular momentum transfer and dissipation often results in both enhanced demagnetization as well as a faster magnetization recovery. We argue that this is exactly what is happening with the non-collinear magnetic regions inside the DWs. The enhanced spin scattering within DWs is a rather old topic born with studies of the extra contribution to the static magnetoresistance [Viret96] or the induced spin transfer torques resulting in their current-induced displacement. To this aim, ballistic models have been developed and can be appropriately adapted for the ultrafast demagnetization scenario in which superdiffusive spin currents play a central role [Battiato10]. The behavior of ballistic spin carriers can be described such as a classical spinned particle perceiving a time varying exchange field while crossing the wall [Viret96, Vanhaverbeke07]. Let us recaller salient features. First, these are band particles that are coupled by exchange to the localized spins (through the so-called *s-d* Hamiltonian). Their velocity perpendicular to the wall is related to their momentum in *k*-space. With the appropriate parameter renormalization, the problem is equivalent to the "fast adiabatic passage" known, e.g., in NMR theory. The spin evolution is given by the Landau-Lifshitz equation:

$$\frac{d\vec{\mu}}{dt} = \frac{J_{ex}S}{\hbar}\vec{m} \times \vec{\mu}$$

where $\vec{\mu}$ is the electron spin, $J_{ex}S$ the exchange energy with the localized moment ($S$) and $\vec{m}$ the direction of the time varying exchange field seen by the ballistic electrons. The localized moments are rotating in a Néel fashion within the DW and the problem is generally treated in this rotating frame [Vanhaverbeke07]. Basically, the electronic spins will precess around the localized moment effective field and thus acquire a component out of the plane of rotation, inducing a torque parallel to the chiral vector: $S_i \times S_j$. The electron spin precession angle ω is proportional to the velocity *v* divided by exchange times and the DW width $2\pi\Delta$ [Viret96]: $<\omega> = \frac{\pi\hbar v}{J_{ex}S\,2\pi\Delta}$. Typically, for electrons at the Fermi level, this precession angle is found to be around 7 degrees for a DW width $2\pi\Delta$ of 15 nm [Vanhaverbeke07]. However, it is to be noticed that this angle can be quite different for the hot electrons produced in the demagnetization process as the relevant parameter values are hard to quantify. Although their velocities should not be too far from those at the Fermi level (in the $10^6$ m/s range [Kampfrath13]), the exchange energies effective in bands over 1 eV above the Fermi level can be dramatically reduced (~ 0.1 eV). Therefore, the expected mistracking angle could be significantly greater for a large part of the hot electrons' distribution. All these processes shall in turn generate a torque applied on the localized moments [Waintal04]. However, because the hot spin currents flow in all directions, mistracking angles can be both positive and negative, resulting in cancellation of the net torque acting on the DWs. The overall effect of the *incoherent* precession results in an average loss of angular momentum. This should speed up the spin relaxation processes within the DW so that after some 100 fs, a net spin current is established from the domains into the interior of the DWs.

The new components of the spin-transfer torque resulting from this latter spin current originating from the *coherent* evolution of the hot electron spins are not cancelled out. Importantly such torques are of opposite sign on the two sides of the DW and should induce a sizeable tilting of the DW magnetization out of the Néel plane as illustrated in Fig. 3(a). This phenomenon is at the origin of a new transient DW shape, made of a Néel type center surrounded by opposite Bloch types as depicted in Fig. 3(b). Such a mixed Bloch/Néel/Bloch contribution will in turn lead to a transient reduction of the measured chirality as it adds two (opposite) Bloch components on both sides of the DW compared to the originally purely Néel character. In order to estimate the amplitude of this DW distortion, it is useful to realize that unlike small current-induced electron flows at the Fermi level, spin fluxes during demagnetization are enormous as for each pulse, typically 0.5 electrons per Co atom are excited to higher bands for the used laser fluence [Kampfrath13]. The timescale for the onset of the induced torques is given by the exchange energy and falls in the 10-fs range, ensuring that the wall distortion does not lag from the population of hot electrons. For a spin temperature sufficiently different between domains and DWs, a quantitative estimate using the abovementioned parameters gives a precession angle of the magnetization inside the DW that is larger than 10 degrees. Moreover, the onset of this Bloch component in the DW must leaks out into the domains, thus slightly increasing the effective DW width as also observed experimentally. The measured expansion of the DW can be directly derived from the variation of the dichroic peak position and width shown in Fig. 2(c). We find that the DW width (slightly) increases rapidly and its magnetization reaches a minimum around 1 ps (blue curve), as reported previously for Bloch type DWs [Pfau2012]. Note that this DW expansion takes place when the quenched magnetization starts to recover (1 ps). After reaching it maximum expansion, the DW width then recovers its original (unpumped represented as dotted lines in Fig. 2(c) size at a timescale of ~5 ps.

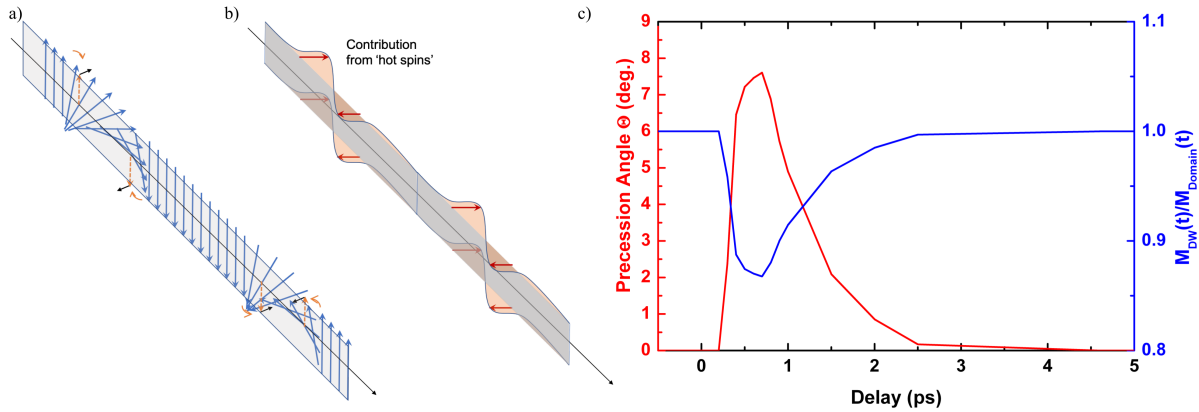

*Figure 3: Magnetization texture modification by hot electrons. (a) Schematic representation of the torque (black arrows) imposed by the 'hot spins' flowing from the domains to the DWs resulting in transient mixed Bloch/Néel/Bloch contributions. (b) Transient DW shape. (c) Precession angles (red) and DW magnetization normalized by Domain one (blue) used in the simulations of the asymmetry ratio shown in Fig. 2(b).*

Using a 1D magnetization profile (described in Supplementary Material S3) and considering the experimental change of magnetization (extracted directly from the square root of the (CL+CR) intensity), the time evolution of the asymmetry ratio can be simulated. We consider a magnetization in the domains extracted from the (CL+CR) data, along with a further 12% reduction of the magnetization inside the DWs to account for incoherent effects, as well as a transient Bloch-Néel-Bloch wall as shown in Fig. 3(a) for coherent ones. With these simulations, we find that the precession angle can reach at the maximum about 8 degrees after a time delay of ~ 0.6 ps [red curve in Fig. 3(c)] simultaneously with the reduction of the DW magnetization [relative to domain magnetization blue curve in Fig. 3(c)], The resulting simulated asymmetry ratio using the described model is plotted as the green curve in Fig. 2(b), and is in excellent agreement with the experimental measurements. Even accounting for DW expansion [red curve in Fig. 2(b)], the agreement can be obtained for a ~10 degrees tilt angle. Although the exchange driven DW distortion is established on a very short timescale, it should last for the nanosecond timescale of the micromagnetic evolution. On the other hand, the incoherent part of the spin current shall relax at the ps timescale of the remagnetization processes, similarly to what we have measured. Interestingly, enhanced spin relaxation existing inside the DWs should speed up remagnetization, explaining that the asymmetry ratio can exceed 1, again in agreement with the experimental results.

In conclusion, we report here about the experimental investigation of the ultra-short timescale evolution of complex chiral Néel spin textures after laser induced demagnetization. Circular dichroism in x-ray resonant magnetic scattering is used to obtain information in the time domain about both the magnetic domain configuration and the magnetic chirality. Beyond the evolution of the period of the magnetic domains in magnetic multilayers with large perpendicular anisotropy, we acquire new insights into the way that the chirality of the non-collinear spin textures, and their long-range ordering, is evolving in the few ps after demagnetization by a strong optical pulse. We observe that the magnetic difference CL-CR (reflecting the DW properties) reduces faster than the diffracted sum signal (associated to domain magnetization) in the first 2 ps after the laser pulse. To explain this unexpected change of XRMS chirality signal at this short timescale, we propose that angular momentum flowing from the interior of the domains inside the DWs associated to hot electrons induces an ultrafast distortion of the DW magnetization. This transient in-plane deformation of the DWs leads to a transient mixed Bloch-Néel-Bloch DW accompanied by an increase of the DWs width and a reduction of the magnetization inside the DW. These original experimental results are reproduced by calculations, considering a magnetization reduction of 12% with a 7.5 degrees distortion of the DW. On a longer timescale, i.e., after a few ps, the DWs return to their pure chiral Néel configuration preserving the original sense of rotation (i.e., chirality) together with a recovery of their magnetization. We emphasize that our approach using dichroism in x-ray resonant scattering is applicable to any other magnetic chiral texture and should provide a better understanding of the evolution of the chirality of spin textures on the ultrafast timescale.

**Acknowledgment:** We acknowledge Ivaylo Nikolov, Michele Manfredda, Luca Giannessi, and Giuseppe Penco for their inestimable help to set up the FEL and the laser for our experiment. NJ would like to thanks S. Flewett for discussion on XRMS simulations. Financial supports from FLAG-ERA SographMEM (ANR-15-GRFL-0005), funding from the Agence Nationale de la Recherche, France, under grant agreement no. ANR-17-CE24-0025 (TOPSKY) and 18-CE24-0018-01 (SANTA), the Horizon2020 Framework Program of the European Commission under FET-Proactive Grant agreement no. 824123 (SKYTOP). E. J. is grateful for financial support received from the CNRS-MOMENTUM. O. S. L., A. V. S., and V. V. K. acknowledge funding from the Engineering and Physical Sciences Research Council (EPSRC) in the United Kingdom (Grant No. EP/T016574/1).